\documentclass[%
superscriptaddress,
nofootinbib,
 amsmath,amssymb,
 aps,
 twocolumn,
 prd
]{revtex4}

\usepackage{amsmath,amssymb}
\usepackage{wasysym}
\usepackage{graphicx}
\usepackage{slashed}
\usepackage{color}
\usepackage[symbol]{footmisc}
\usepackage{ulem}
\newcommand{\beq}{\begin{equation}}
\newcommand{\eeq}{\end{equation}}
\newcommand{\bea}{\begin{eqnarray}}
\newcommand{\eea}{\end{eqnarray}}

\allowdisplaybreaks
\begin{document}
\author{Shou-Shan Bao}
\email{ssbao@sdu.edu.cn}
\affiliation{Institute of Frontier and Interdisciplinary Science,
Key Laboratory of Particle Physics and Particle Irradiation (MOE),
Shandong University, Qingdao 266237, China}
\author{Qi-Xuan Xu}
\email{qixuan.xu@student.manchester.ac.uk}
\affiliation{Department of Physics and Astronomy, The University of Manchester, Oxford Road, Manchester
M13 9PL, UK}
\affiliation{College of Science, China University of Petroleum, Qingdao, 266580, China}
\author{Hong Zhang}
\email{hong.zhang@sdu.edu.cn}
\affiliation{Institute of Frontier and Interdisciplinary Science,
Key Laboratory of Particle Physics and Particle Irradiation (MOE),
Shandong University, (QingDao), Shandong 266237, China}
\date{\today}

\title{Improved Analytic Solution of Black Hole Superradiance}

\begin{abstract}
The approximate solution of the Klein-Gordon equation for a real scalar field of mass $\mu$ in the geometry of a Kerr black hole obtained by Detweiler \cite{Detweiler:1980uk} is widely used in the analysis of the stability of black holes as well as the search of axion-like particles. In this work, we confirm a missing factor $1/2$ in this solution, which was first identified in Ref.~\cite{Pani:2012bp}. The corrected result has strange features that put questions on the power-counting strategy. We solve this problem by adding the next-to-leading order (NLO) contribution.  Compared to the numerical results, the NLO solution reduces the percentage error of the LO solution by a factor of 2 for all important values of $r_g \mu$. Especially the percentage error is $\lesssim 10\%$ in the region of $r_g\mu \lesssim 0.35$. The NLO solution also has a compact form and could be used straightforwardly.
\end{abstract}
\maketitle
%

\section{Introduction}

If a light scalar boson exists with a proper value of mass, it could form gravitational bound states around spinning black holes (BHs). The bound states can continuously extract energy and angular momentum from the host BHs until the nonlinear effect is important or the angular momentum of the BH is below some critical value. This phenomenon is often referred to as superradiance, which has been applied in many research frontiers, including the stability of spinning BHs \cite{Huang:2018qdl,Boskovic:2018lkj,Franzin:2021kvj,Garcia-Saenz:2021uyv,Lin:2021ssw, Biswas:2021gvq,Guo:2021xao, Herdeiro:2021znw} and the search of axion-like-particles (ALPs) \cite{Arvanitaki:2010sy,Hui:2021tkt}. ALP is one of the most popular candidates of dark matter in our universe, with the mass ranging from $10^{-22}$~eV to a few eV. Especially, the formation of the ALP clouds from superradiance only depends on the ALP mass, not on its couplings to the Standard Model particles, making the study of superradiance a model-independent way to search for light dark matters. Many observational strategies of superradiance have been proposed, such as the BH shadow \cite{Roy:2019esk,Creci:2020mfg,Davoudiasl:2019nlo,Cunha:2019ikd,Roy:2021uye}, the birefringent effect of the lights traversing through the boson clouds \cite{ Yuan:2020xui, Chen:2019fsq, Cannizzaro:2021zbp, Liu:2021llm}, and the gravitation wave signals generated by the spinning boson clouds around the host BHs \cite{Arvanitaki:2014wva, Arvanitaki:2016qwi, Baryakhtar:2017ngi, Brito:2017wnc, Brito:2017zvb, Hannuksela:2018izj, Isi:2018pzk, Palomba:2019vxe, Sun:2019mqb, Brito:2020lup, Barausse:2020rsu, Zhu:2020tht, Ng:2020jqd, Baryakhtar:2020gao, Aggarwal:2020olq, Chen:2021lvo, Yuan:2021ebu, Ghosh:2021uqw, Berlin:2021txa}. It is also believed that the superradiant boson clouds could modify the gravitational waveform of two-BH-merger events \cite{Ghosh:2018gaw,Baumann:2018vus,Zhang:2019eid,Hannuksela:2019vip,Annulli:2020lyc,Ikeda:2020xvt, Choudhary:2020pxy, Tong:2021whq, DeLuca:2021ite, Chung:2021roh,Payne:2021ahy,Su:2021dwz,Takahashi:2021yhy}. For more interesting work with superradiance, we refer the readers to the recent review \cite{Brito:2015oca}.

All of these studies rely on the calculation of the boson bound states. Due to the superradiance, the eigen-frequency is a complex number \cite{Vishveshwara:1970zz}. The direct numerical calculation requires a 2-dimensional shooting algorithm. Very high numerical precision has to be kept, because the imaginary part of the eigenfrequency is at least 7 orders of magnitude smaller than the real part. Until now, no success has been achieved in this direction. With the indirect method proposed firstly by Leaver \cite{Leaver:1985ax},  the eigenfrequencies of the lowest several partial waves are successfully calculated with the principal number $n$ fixed to be zero \cite{Cardoso:2005vk,Dolan:2007mj}. This numerical calculation still needs very high precision, which is nontrivial to be reproduced. Besides the numerical solution, approximate analytic expressions also exist. If the multiplication of masses of the host BH and the ALP is much less than 1, a beautiful analytic approximation has been proposed by Detweiler \cite{Detweiler:1980uk}. It has a compact form and has been widely used in literatures. Another approximation based on the WKB approximation is also available \cite{Zouros:1979iw}. However, these three solutions do not agree with each other, with differences of more than 100\% in the regions where the approximations are expected to be valid. This raises the question about which solution is correct, or none of them is. Without solving this puzzle, most of the efforts on superradiance stop at qualitative descriptions or order-of-magnitude estimates.

In this work, we solve the puzzle by carefully investigating the leading-order (LO) approximation made by Detweiler \cite{Detweiler:1980uk}. We find a missing factor $1/2$ in the treatment of infinities, which was also discussed at the end of Ref.~\cite{Pani:2012bp}. Here we confirm their finding. The corrected result has a better agreement with the numerical solution, but they still do not converge in the region where the approximate solution is expected to be excellent. By careful study of the power-counting, we find a next-to-leading order (NLO) term which is enhanced by $(r_g^2-a^2)^{-1/2}$, thus has an important effect. After systematically adding the NLO contribution, the improved approximation agrees much better with the numerical results. More importantly, the improved solution converges to the numerical result in the region where the approximation is expected to work well, qualifying our strategy of power-counting. The improved solution has a compact form and can be applied straightforwardly in future studies of superradiance.

In the rest of this article, we first review the previous calculation in Sec.~\ref{sec:review} and point out a possible reason for the additional factor $1/2$. It also sets the stage for the NLO correction. Then in Sec.~\ref{sec:NLO} the power-counting is argued and the NLO contribution is added. Comparisons with the numerical calculation are also provided. Finally, a summary is given in Sec.~\ref{sec:summary}. We choose $\hbar=c=1$ throughout this article.

\section{Review of LO Calculation}\label{sec:review}

A real spin-0 boson with mass $\mu$ can be described by a real scalar field $\phi(x)$. The action for $\phi$ and the space-time metric tensor $g_{\mu\nu}(x)$ in general relativity is
\begin{align}\label{eq:action}
S=\int d^4 x \sqrt{-g}\, \left[\frac{1}{2}\,g^{\mu\nu} \nabla_\mu \phi \, \nabla_\nu \phi - \mathcal{V}(\phi)+\frac{R}{16\pi G}\right],
\end{align}
where $g^{\mu\nu}$ is the inverse of the metric tensor, $g$ is its determinant, $\mathcal{V}(\phi)$ is the potential energy density of the scalar field, $R$ is the space-time curvature scalar, and $G$ is Newton's gravitational constant. We use a metric with signature $(+1,-1,-1,-1)$. Varying the action with respect to the scalar field gives the Klein-Gordon equation in curved spacetime, 
\begin{align}\label{eq:KG-g}
g^{\mu\nu}\nabla_\mu \nabla_\nu \phi + \frac{\delta\mathcal{V}}{\delta\phi} = 0.
\end{align}
Varying $S$ with respect to the metric gives the Einstein equation, in which the stress tensor is from the real scalar field $\phi$.

The coupled Einstein equation and the Klein-Gordon equations are very difficult to solve, even numerically. Perturbative methods have been employed to simplify the calculation. For all physically interesting axion models, the self-interaction is always suppressed by the axion decay constant $f_a$, which is around $10^{11}$~GeV for the QCD axion and can be even higher for axion-like-particle models. Thus the self-interaction of axions can be considered as a perturbation. Moreover, since the superradiance is relatively slow and the nonlinear effects are expected to terminate the process before the cloud accumulates too many bosons, the modification of the axion cloud to the Kerr space-time metric is also small. By taking both the axion self-interaction and the effect of the axion cloud on the Kerr metric as perturbations, the problem is reduced to a Klein-Gordon equation for a free real scalar field on the static Kerr background.

In this work, we use the Boyer-Lindquist coordinates \cite{Boyer:1966qh}. The solution of a Kerr BH with spin $J$ and mass $M$ has the line element in the form,
\begin{align}\label{eq-Kerr-ds2}
\begin{split}
ds^2 =& \left( 1-\frac{2\, r_g r}{\Sigma} \right) dt^2
+ \frac{4\,a\, r_g r}{\Sigma}\sin^2\theta \, dt \, d\varphi
- \frac{\Sigma}{\Delta} dr^2\\
&\hspace{-0.5cm}
-\Sigma\, d\theta^2 -\left[(r^2+a^2)\sin^2\theta + 2\frac{r_g r}{\Sigma}a^2\sin^4\theta\right] d\varphi^2.
\end{split}
\end{align}
where $a = J/M$, $r_g = G\, M$, $\Delta = r^2-2r_g r + a^2$, and $\Sigma = r^2 + a^2 \cos^2\theta$. The equation $\Delta=0$ gives two event horizons at $r_\pm = r_g \pm b$ with $b=(r_g^2-a^2)^{1/2}$.

For real scalars with no self-interaction, the potential $\mathcal{V}(\phi)$ has only the mass term $\mathcal{V}(\phi) = \mu^2 \phi^2/2$. Insert $g_{\mu\nu}$ from Eq.~\eqref{eq-Kerr-ds2} into the Klein-Gordon equation in Eq.~\eqref{eq:KG-g}, we can obtain the equation of motion for a real scalar on the Kerr metric. Surprisingly, the variables of the field can be separated in the form of \cite{Press:1973zz},
\begin{align}
\phi(t,\vec{r})
=\sum_{l,m} \int d\omega  \left[ e^{i(m\varphi-\omega t)} R_{lm}(r) S_{lm}(\theta) +\text{c.c.}\right].
\end{align}
The equations for the radial and angular wave functions are
\begin{subequations}
\begin{align}
\begin{split}\label{eq:radial-R}
&\Delta\frac{d}{d r}\left(\Delta \frac{d R_{lm}}{d r}\right)
+ \Big[ \omega^2 (r^2+a^2)^2 - 4\,a\, r_g  r\, m \,\omega \\
&\hspace{0.8cm}
 + a^2 m^2 - (\mu^2 r^2 + a^2 \omega^2 + \Lambda_{lm}) \Delta \Big] R_{lm} = 0,
\end{split}\\
\begin{split}\label{eq:angular}
&\frac{1}{\sin\theta}\frac{d}{d \theta}\left(\sin\theta \frac{d S_{lm}}{d \theta}\right)
+\Big[ -a^2 \kappa^2\cos^2\theta \\
& \hspace{3cm}
-\frac{m^2}{\sin^2\theta} +\Lambda_{lm}\Big] S_{lm} = 0,\\
~
\end{split}
\end{align}
\end{subequations}
where $\kappa=\sqrt{\mu^2-\omega^2}$ and $\Lambda_{lm}$ is the eigenvalue of the angular equation.

The solution to Eq.~\eqref{eq:angular} is the spheroidal harmonics, with $\Lambda_{lm}$ being the eigenvalue~\cite{Berti:2005gp}. The challenge lies in solving the radial eigen-equation in Eq.~\eqref{eq:radial-R}, with the wavefunction approaching zero at infinity. The eigenfrequency $\omega$ is a complex number and the numerical method requires a 2-dimensional shooting algorithm. In addition, the imaginary part of $\omega$ is orders of magnitude smaller than its real part,  requiring very high precision in the numerical calculation. Lacking accurate results restricts the development in topics related to superradiance, with many efforts stopping at qualitative descriptions or order-of-magnitude estimates.

In the limit of small $r_g\mu$, Detweiler proposed a beautiful method calculating the complex eigenfrequency $\omega$ ~\cite{Detweiler:1980uk}.  Below is a review of this method. We introduce a power-counting parameter $\alpha \sim r_g \mu$ for the expansion. The scaling of other parameters are $\text{Re}\,\omega \sim \mu$ and $a\sim r_+\sim r_- \sim r_g$. At the $r\gg r_g$ limit, the radial equation is,
\begin{align}\label{eq:R-2}
\begin{split}
\frac{d^2}{d r^2}\left(rR\right) &
+ \Big[-\kappa^2
 + \frac{2\kappa \lambda}{r}- \frac{l'(l'+1)}{ r^2} \Big] (rR) = 0,
\end{split}
\end{align}
where $l'=l+\epsilon$ and,
\begin{align}
 \lambda= r_g (2\omega^2-\mu^2)/\kappa. \label{eq:lambda}
\end{align}
Here $\epsilon \sim \mathcal{O}(\alpha^2)$ plays the role of a regulator. Its value is unimportant for LO calculation. Nonetheless, it cannot be trivially dropped. In Ref.~\cite{Detweiler:1980uk}, $\epsilon$ was set to zero and $l'=l$ from the beginning, which leads to a factor of $1/2$ missing in the final result, as will be clear soon. The bound state wave function decays exponentially at infinity. Up to an arbitrary normalization, the solution can be written in terms of the confluent hypergeometric function,
\begin{equation}\label{eq:ana-R-large-r}
R(r)= e^{-\kappa r}(2\kappa r)^{l'}  U(l'+1-\lambda,2\,l'+2;2\,\kappa\, r).
\end{equation}

In the small $r$ region, the radial function can be written in terms of  $z=(r-r_+)/2b$, 
\begin{align}
z(z+1)\frac{d}{dz}\left[ z(z+1)\frac{dR}{dz}\right]
+V(z) R = 0,
\end{align}
where $V(z)$ is a polynomial of $z$,
\begin{widetext}
\begin{align}\label{eq:small-r-V}
\begin{split}
V(z) &= p^2
+\left[4 b^{-1} r_g r_+ \omega (r_+ \omega - r_g \omega_c) -(\Lambda_{lm}+r_+^2\mu^2 + a^2 \omega^2)\right]z
+ \left(a^2\omega^2 -\Lambda_{lm} + 2\mu^2 a^2 -3\mu^2 r_+^2 +6 r_+^2 \omega^2\right)z^2\\
&
+4b \left[r_g \mu^2 + 2 r_+ (\omega^2 - \mu^2)\right] z^3
-4 b^2\kappa^2 z^4,
\end{split}
\end{align}
\end{widetext}
with $p={r_g r_+}(\omega- \omega_c)/b$ and $\omega_c=a m/2r_g r_+$, both of which scale as $\alpha$. At LO of $\alpha$, the $V(z)$ is $p^2 - l'(l'+1)z(z+1)$ and the solution is proportional to Gauss hypergeometric function. Changing the variable from $z$ back to $r$, the solution is,
\begin{equation}\label{eq:ana-R-small-alpha}
R(r) = \left(\frac{r-r_+}{r-r_-}\right)^{-ip} \!\!\!\!\!
_2F_1\left(-l',l'+1;1-2ip;-\frac{r-r_+}{2b}\right),
\end{equation}
up to an arbitrary normalization.

The solution in Eq.~\eqref{eq:ana-R-large-r} is valid when $r\gg r_g$. The solution in Eq.~\eqref{eq:ana-R-small-alpha} requires $r\ll r_g \alpha^{-2}$ from the ignorance of terms proportional to $z^3$ and $z^4$. The two solutions have an overlap region when $\alpha\ll 1$. In Ref.~\cite{Detweiler:1980uk}, the author took the small $r$ limit of Eq.~\eqref{eq:ana-R-large-r}, which is,
\begin{equation}\label{eq:ana-R-large-r-small-r-limit}
\frac{(2 \kappa )^{l'}\Gamma(-2l'-1)}{\Gamma(-l'-\lambda)} r^{l'}+
\frac{(2\kappa)^{-l'-1} \Gamma(2l'+1)}{\Gamma(l'+1-\lambda)} r^{-l'-1},
\end{equation}
and the large $r$ limit of Eq.~\eqref{eq:ana-R-small-alpha}, which is,
\begin{align}
\begin{split}
\frac{(2b)^{-l'} \Gamma(2l'+1) }{\Gamma(l'+1)\Gamma(l'+1-2ip)} r^{l^\prime} 
+
\frac{(2b)^{l'+1} \Gamma(-2l'-1)}{\Gamma(-l'-2ip) \Gamma(-l')}r^{-l'-1}.
\end{split}
\end{align}
In the overlapped region, the ratio of the coefficients of the $r^{l'}$ and $r^{-l'-1}$ must be the same for the two solutions.
The obtained equation can be solved numerically for $\omega$. It can also be solved perturbatively with the observation that the coefficient of $r^{-l'-1}$ in expression~\eqref{eq:ana-R-large-r-small-r-limit} must be severely suppressed for the wavefunction to be convergent at small $r$. It means $l'+1-\lambda$ is in the neighbourhood of zero or some negative integer,
\begin{align}\label{eq-delta-lambda}
l'+1-\lambda = -n-\delta\lambda, \text{ with } n\geq 0.
\end{align}
Combining this equation with Eq.~\eqref{eq:lambda} gives $r_g \kappa\sim \mathcal{O}(\alpha^2)$. The equation of the ratio of the coefficients can then be solved to the LO of $\delta\lambda$. In Ref.~\cite{Detweiler:1980uk}, the regulator $\epsilon$ was taken to be zero from the very beginning. The resulted ill-defined piece $\Gamma(-2l-1)/\Gamma(-l)$ then has to be treated with great caution. We conjecture this ratio was mistakenly replaced by $(-1)^{l+1}l!/(2l+1)!$ in Ref.~\cite{Detweiler:1980uk}. If considering the regulator correctly by $l'=l+\epsilon$ and taking $\epsilon \to 0$ at the end, one obtains an additional factor of $1/2$. The corrected result is,
\begin{align}\label{eq-delta-lambda-improve}
\begin{split}
\delta\lambda^{(0)} =& -ip \,(4\kappa b)^{2l+1}
\frac{(n+2l+1)!(l!)^2}{n!\left[(2l)!(2l+1)!\right]^2} \prod_{j=1}^l (j^2+4p^2),
\end{split}
\end{align}
which scales as $\mathcal{O}(\alpha^{4l+3})$ and the superscript (0) indicates that it is the LO contribution of the imaginary part of $\omega$. This correct $\delta\lambda^{(0)}$ was also obtained at the end of Ref.~\cite{Pani:2012bp}. Here we confirm their result. One could also get this result without using the regulator. 
The subtle point is using the correct identities which are valid for $\Gamma$ functions with negative integer arguments.
In comparison, the calculation with the regulator $\epsilon$ is straightforward.
More details are explained in the appendix. Defining $\omega = \omega_0 + \omega_1 \delta\lambda^{(0)}$, using the definition of $\lambda$ in Eq.~\eqref{eq:lambda} and $\delta\lambda$ in Eq.~\eqref{eq-delta-lambda}, one could obtain $\omega_0$ and $\omega_1$ with $\epsilon \to 0$,
\begin{subequations}\label{eq:omegas}
\begin{align}
\omega_0 &= \mu\left( 1- \frac{2 r_g^2 \mu^2}{\bar{n}^2 + 4r_g^2 \mu^2 +\bar{n}\sqrt{\bar{n}^2+8 r_g^2 \mu^2}}\right)^{1/2},\\
\omega_1 
&=\frac{\mu^2-\omega_0^2}{\bar{n}\,\omega_0} 
\left[1 + \frac{4 r_g^2}{\bar{n}^2}(2\omega_0^2 - \mu^2)\right]^{-1},
\end{align}
\end{subequations}
where $\bar{n} = n+l+1$. Eqs.~\eqref{eq-delta-lambda-improve} and \eqref{eq:omegas} give the LO approximation of $\omega$.

To judge how good the corrected solution compared to the numerical result, we follow the continued fraction method, which is firstly proposed by Leaver \cite{Leaver:1985ax} and developed in Refs.~\cite{Cardoso:2005vk,Dolan:2007mj}. The radial function is firstly expanded as an infinite power series,
\begin{align}
R(r)=(r-r_+)^{-ip}(r-r_-)^{ip+\lambda-1}e^{-\kappa r} \sum_{n=0}^{+\infty}a_n\left(\frac{r-r_+}{r-r_-}\right)^n.
\end{align}
Inserting it into Eq.~\eqref{eq:radial-R}, one could obtain a 3-term recursive relation. This relation can be rewritten into a continued fraction, which relate the ratio of two successive coefficients $a_{n+1}/a_n$ to $a_1/a_0$. Requiring the wave function $R(r)$ decays at large $r$, one could get another expression for $a_{n+1}/a_n$ at $n$ approaching infinity. Combining these two expressions for the ratio, an ``eigen-equation" of $\omega$ written in terms of continued fraction is thus obtained. Solving this equation numerically requires high precision because the imaginary part of the eigenvalue $\omega$ is at least seven decades smaller than its real part. With a self-written code, we could obtain the numerical solution for different values of $n$, $l$, and $m$ with numerical errors less than $10^{-7}$. The $n=0$ results agree with the numbers in Ref.~\cite{Dolan:2007mj} with high precision.

\begin{figure}[htbp]
\begin{center}
\includegraphics[clip, trim={0 0.15cm 0 0cm}, width=8.5cm]{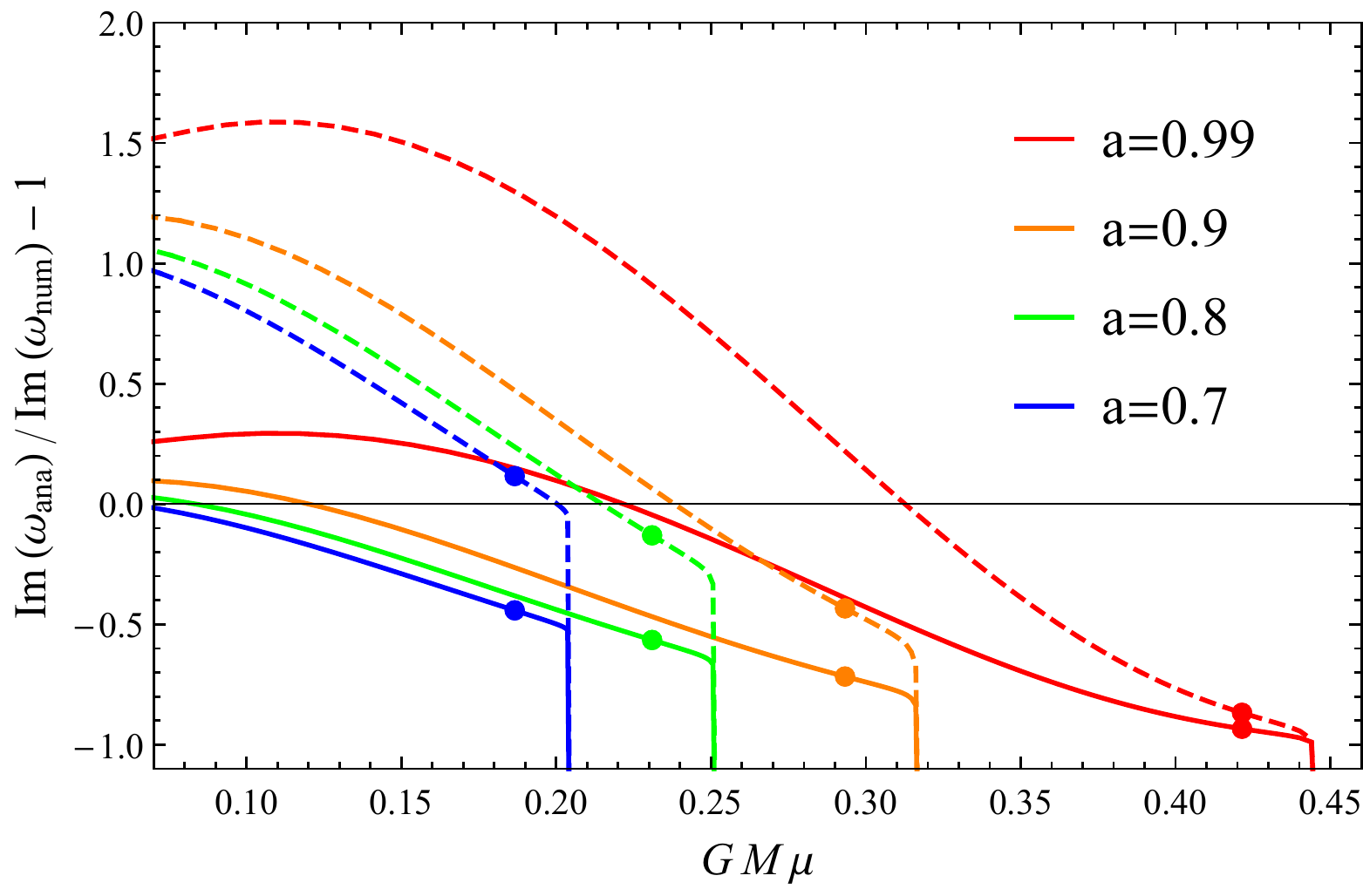}
\end{center}
\caption{Comparison of the numerical result and the analytic approximation for $n=0$, $l=m=1$. The solid curves and the dashed curves are from Eq.~\eqref{eq-delta-lambda-improve} and Ref.~\cite{Detweiler:1980uk}, respectively. The round dot on each curve labels the position of largest $\text{Im}(\omega_\text{num})$ for each $a$. The numerical values are calculated with the method in Ref.~\cite{Dolan:2007mj}.}\label{fig:err-Detweiler}
\end{figure}

In Fig.~\ref{fig:err-Detweiler}, we show the percentage errors of $\text{Im}( \omega)$ comparing to the numerical results, with $n=0$, $l=m=1$. We compare our results in Eq.~\eqref{eq-delta-lambda-improve} and the previous results from Ref.~\cite{Detweiler:1980uk}. Previous analytic solutions have percentage errors of around $150\%$ at small $r_g\mu$, while the corrected solutions reduce the errors to about $40\%$. This improvement, however, is still not satisfactory. For very small $r_g \mu$, where the analytic approximation is supposed to work well, the error is at first a constant as much as $30\%$, then cross the horizontal axis from above. Another strange feature is that the errors at small $r_g\mu$ increase with $a$.

We go back to Eq.~\eqref{eq:small-r-V} and its LO approximation to understand these behaviours. To obtain the LO approximation from Eq.~\eqref{eq:small-r-V}, we implicitly assume $\alpha\ll (b/r_g)^{1/2}$ from ignoring the first term in the coefficient of $z$. For a fast-rotating BH, the value of $b$ is very small and this assumption is satisfied only for very small values of $\alpha$. It explains why the analytic and numerical results do not agree even with $r_g \mu$ as small as 0.1, as well as larger error for larger $a$.

\section{Next-to-leading order correction}\label{sec:NLO}

To avoid the restriction of this assumption, it is crucial to include the first term in the coefficient of $z$ in Eq.~\eqref{eq:small-r-V}. The NLO correction of $\alpha$ is also added without sacrificing the compactness of the result. We first solve the regulator $\epsilon$ explicitly. From Eq.~\eqref{eq:radial-R} and the expanded form of $\Lambda_{lm}$, one could obtain,
\begin{align}
\epsilon =- \frac{8}{2l+1}(r_g \mu)^2+\mathcal{O}(\alpha^4).
\end{align}
We also write the coefficient of $z$ in Eq.~\eqref{eq:small-r-V} as {$-l'(l'+1)+q$, where $q$ is defined as,
\begin{align}
q=4\,r_g \,\omega\, p-2(4\,r_g-r_+)\,r_g\mu^2+\mathcal{O}(\alpha^4).
\end{align}
Even with the presence of $q$, the equation can still be solved with a compact solution. Up to an arbitrary normalization, the corresponding radial function is then,
\begin{align}\label{eq:ana-R-small-alpha-2}
\begin{split}
R(r) =& \frac{(r-r_-)^{\sqrt{q-p^2}}}{(r-r_+)^{ ip}}
\,_2F_1\Big( -l' - ip + \sqrt{q-p^2}, \\
& l'+1 -  ip +\sqrt{q-p^2}; 1 - 2ip;-\frac{r-r_+}{2b}\Big).\\
&~
\end{split}
\end{align}
In the $r\to +\infty$ limit, this function behaves as
\begin{align}
\begin{split}
 &\frac{(2b)^{-l'- ip +\sqrt{q-p^2}} \Gamma(2l'+1)\Gamma(1-2ip)}{\Gamma(l'+1-ip-\sqrt{q-p^2})\Gamma(l'+1-ip +\sqrt{q-p^2})} r^{l'}\\
 &+
\frac{(2b)^{l'+1- ip +\sqrt{q-p^2}} \Gamma(-2l'-1)\Gamma(1-2ip)}{\Gamma(-l'-ip-\sqrt{q-p^2})\Gamma(-l'-ip+\sqrt{q-p^2})}r^{-l'-1}.
\end{split}
\end{align}
Following similar matching steps, one could obtain the $\delta\lambda$ at NLO after some algebra,
\begin{align}\label{eq:delta-lambda-best}
\delta\lambda^{(1)} = \left(\frac{q}{2\epsilon} -\frac{\epsilon}{2}- ip\right) 
\frac{\left(4\kappa b\right)^{2l'+1}\Gamma(n+2l'+2)\Gamma_{pq}}{n!\left[\Gamma(2l'+1)\Gamma(2l'+2)\right]^2},
\end{align}
where the superscript (1) indicates that it is the NLO result. The $\Gamma_{pq}$ is defined as,
\begin{widetext}
\begin{align}
\Gamma_{pq}
=
\frac{\left| \Gamma(l'+1+ip+\sqrt{q-p^2})\Gamma(l'+1+ip-\sqrt{q-p^2})\right|^2 \Gamma(1+2\epsilon) \Gamma(1-2\epsilon)}{\Gamma(1-ip-\sqrt{q-p^2}-\epsilon)\Gamma(1+ip+\sqrt{q-p^2}+\epsilon)
\Gamma(1-ip+\sqrt{q-p^2}-\epsilon)\Gamma(1+ip-\sqrt{q-p^2}+\epsilon)}
\end{align}
\end{widetext}
This is our major result. Finally, $\omega$ is calculated with the definition of $\lambda$ and Eq.~\eqref{eq-delta-lambda}, 
\begin{align}
\frac{r_g(2\omega^2-\mu^2)}{\sqrt{\mu^2-\omega^2}} = \bar{n} +(\delta\lambda^{(1)}+\epsilon).
\end{align}
One could define $\omega=\omega_0+\omega_1 \delta\lambda^{(1)}$ and solve for $\omega_0$ and $\omega_1$ perturbatively. It turns out to be nontrivial. At NLO, we could equally define,
\begin{align}\label{eq:omega}
\omega = \omega_0+\omega_1(\epsilon +\delta\lambda^{(1)}) + \mathcal{O}(\epsilon^2),
\end{align}
then the obtained $\omega_0$ and $\omega_1$ are the same as in Eqs.~\eqref{eq:omegas}

\begin{figure}
\begin{center}
\includegraphics[clip,trim={0 0.2cm 0 0} ,width=8.5cm]{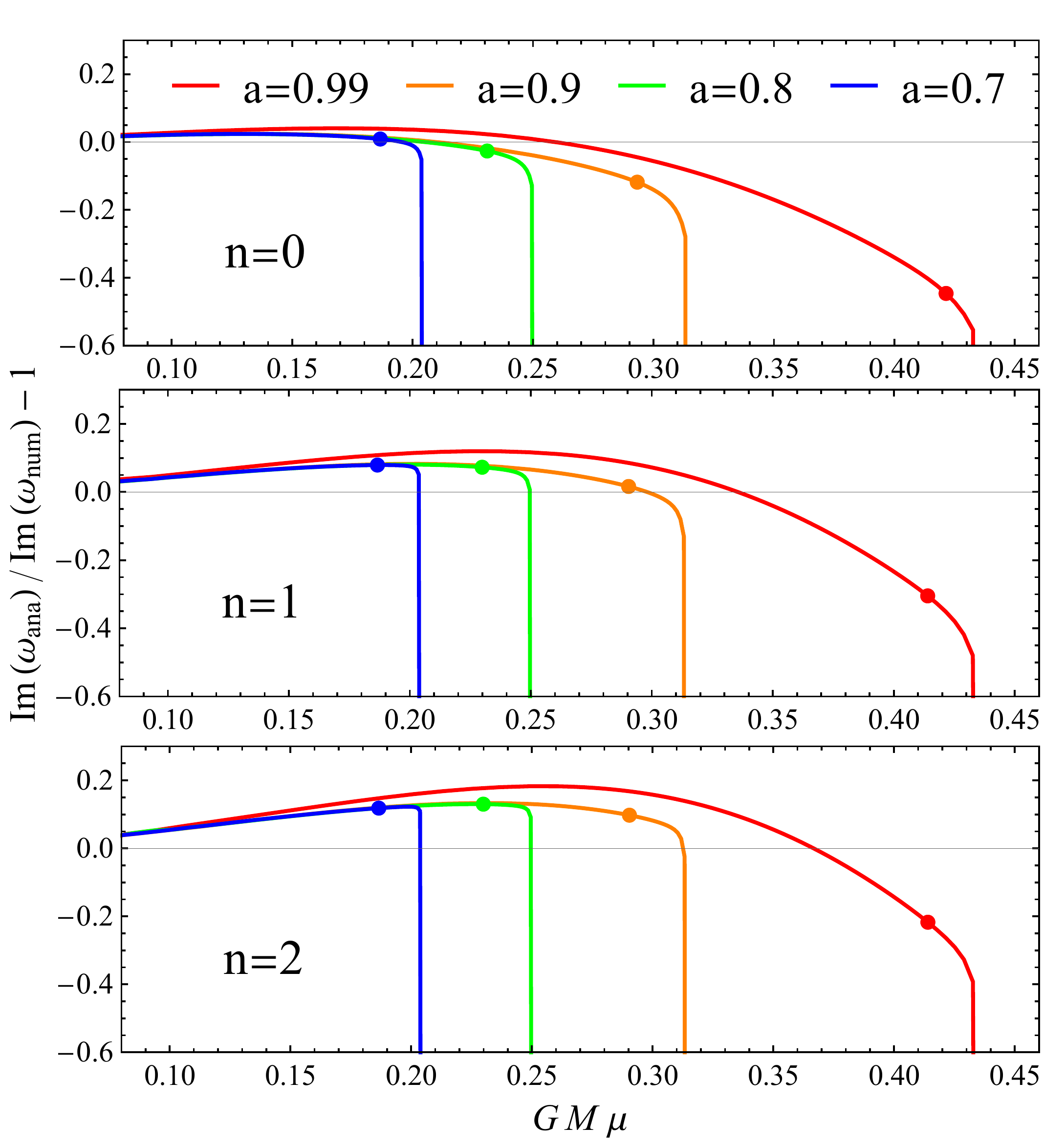}
\end{center}
\caption{Comparison of the numerical result and the improved analytic approximation in Eq.~\eqref{eq:omega} for $l=m=1$ and $n=0,1,2$. The round dot on each curve labels the position of largest $\text{Im}(\omega_\text{num})$ for each $a$.}\label{fig:err-lm-1}
\end{figure} 

\begin{figure}
\begin{center}
\includegraphics[clip,trim={0 2.3cm 0 0} ,width=8.5cm]{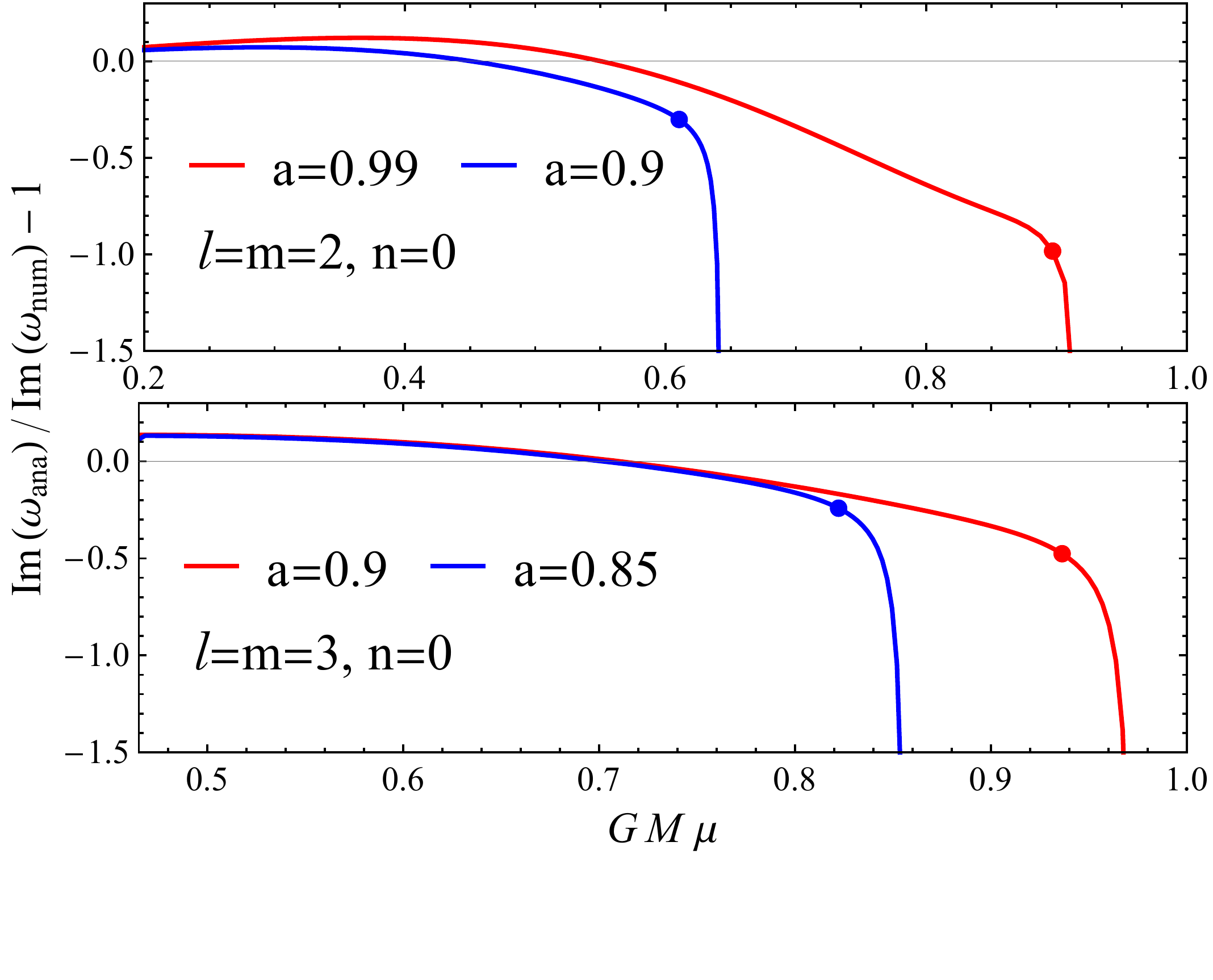}
\end{center}
\caption{Comparison of the numerical result and the improved analytic approximation in Eq.~\eqref{eq:omega} for $n=0$ with $l=m=2$ (upper panel) and $l=m=3$ (lower panel). The round dot on each curve labels the position of largest $\text{Im}(\omega_\text{num})$ for each $a$.}\label{fig:err-lm-23}
\end{figure} 

In Fig.~\ref{fig:err-lm-1}, we show the comparison of this analytic result with the numerical solution for $l=m=1$. For each curve, the error at the point with the largest $\text{Im}(\omega_{\text{num}})$ (labelled with a dot in the figure) is less than 15\% except for very large $a$. The divergences at the right end of the curves are due to the fast dropping of $\text{Im}(\omega_{\text{num}})$ from the maxima to zero (see Fig.~7 of Ref.~\cite{Dolan:2007mj}). Since the regions on the right of the maxima are unimportant for all known physical applications, we safely conclude that the improved analytic approximation for $l=m=1$ is valid with an error less than $30\%$ for all values of $r_g\mu$ and $a$. The similar comparisons for $l=m=2$ and $l=m=3$ are shown in Fig.~\ref{fig:err-lm-23}. The errors are less than $30\%$ for  $a\lesssim 0.9$. Accurate calculations with larger values of $a$ have to rely on the complicated numerical algorithms, such as the one explained in Ref.~\cite{Dolan:2007mj}.

All curves in Figs.~\ref{fig:err-lm-1} and \ref{fig:err-lm-23} gradually deviate from zero when $r_g\mu$ increases. This behavior is expected since the analytic approximation is a truncated Taylor expansion of the exact solution at $\alpha=0$. In getting the small-$r$ solution in Eq.~\eqref{eq:ana-R-small-alpha-2}, keeping $q$ in the calculation removes the restriction from the implicit assumption at LO, such that the small quantity $b$ does not mix with the power counting of $\alpha$ anymore. Note $\Gamma_{pq}$ in $\delta \lambda^{(1)}$ scales as $1+i\epsilon p$, hence the contribution of $q/2\epsilon$ to the imaginary part of $\delta\lambda^{(1)}$ is $\sim iqp$, which is NLO in $\alpha$ compared to $ip$. Therefore the improved analytic approximation is more accurate than the LO result.

Further improvement to higher orders is straightforward, although not very necessary for the current precision requirement. The expression of $\delta \lambda^{(1)}$ in Eq.~\eqref{eq:delta-lambda-best} is valid independent on the truncation of $p$, $q$ and $\epsilon$. For higher orders of $\text{Im}\,\omega$, one only needs to keep higher orders of $\alpha$ in these quantities, as well as keep more terms in Eq.~\eqref{eq:omega}.

\section{Summary}\label{sec:summary}

The solution of the Klein-Gordon equation for a real scalar field of mass $\mu$ in the geometry of a Kerr BH is important in the analysis of the stability of the BH as well as the searching for ALPs. Since the numerical calculation is nontrivial to obtain, the analytic approximation obtained by Detweiler in Ref.~\cite{Detweiler:1980uk} is widely used. The original solution missed an overall $1/2$ which was firstly identified in Ref.~\cite{Pani:2012bp}. In this work, we confirm this extra factor. By comparing the corrected solution with the numerical result obtained with the continued fraction method (see Fig.~\ref{fig:err-Detweiler}), We find the corrected solution agrees better with the numerical result. But it is not satisfying in two aspects. Firstly,  at very small $r_g \mu$ where the analytic approximation is supposed to work well, the percentage error is at first a constant as much as $30\%$, then cross the horizontal axis from above. Secondly, the percentage error at small $r_g\mu$ increases with the BH spin parameter $a$.

After a careful analysis, we find there is a NLO term which is enhanced by a factor of $1/b = (r_g^2-a^2)^{-1/2}$. The ignorance of this term restricts the validity of the LO result to be $\alpha\ll b/r_g$, which is a very small number for fast-spinning BHs. We added the contribution of this term and other NLO terms of order $\alpha$ for the consistency of power-counting. The full NLO solution is also in a compact form and could be used straightforwardly.

By comparing the NLO solution to the numerical result, we find they agree to each other by a percentage error less than $5\%$ at small $\alpha$ for $n=0$ and $l=m=1$ with different values of $a$ (see Fig.~\ref{fig:err-lm-1}). Most importantly, the error decreases to zero at small $\alpha$, validating the power-counting strategy used in the analysis. The percentage error increases for larger numbers of $n$, $l$ and $m$, which is about $10\%$ in the region of $\alpha \lesssim 0.35$ (see Figs.~\ref{fig:err-lm-1} and \ref{fig:err-lm-23}).

\appendix

\section{Calculation of $\Gamma(-2l-1)/\Gamma(-l)$}

In this appendix, we explain in detail the correct way of calculating this ratio with and without the regulator.

The calculation with the regulator is straightforward since both $\Gamma$ functions are well defined.
One could safely use $\Gamma(1+z) = z\Gamma(z)$ repeatedly and get,
\begin{align}
\begin{split}
&\lim_{\epsilon\to 0}\frac{\Gamma(-2l-1-2\epsilon)}{\Gamma(-l-\epsilon)}\\
=&\lim_{\epsilon\to 0}\frac{(-l-\epsilon)\dots(-\epsilon)\Gamma(1-2\epsilon)}{(-2l-1-2\epsilon)\dots(-2\epsilon)\Gamma(1-\epsilon)}\\
=&
\frac{(-1)^{l+1} l!}{2(2l+1)!}.
\end{split}
\end{align}

This result could also be obtained without the regulator.
The following steps are provided by an anonymous referee and we list them here to show the readers a different way of doing the calculation. 
The key formula is 
\begin{align}
\Gamma\left(-\frac{n}{2}\right) =
\frac{(-1)^\frac{n+1}{2} 2^n \sqrt{\pi}}{n!}\left(\frac{n-1}{2}\right)!,
\end{align}
which is valid when $n$ is a positive odd integer.
We will also use 
$\Gamma(z)\Gamma(z+{1}/{2}) = 2^{1-2z}\sqrt{\pi}\,\Gamma(2z)$.
Then
\begin{align}
\begin{split}
\frac{\Gamma(-2l-1)}{\Gamma(-l)}
&= \frac{1}{(-2l-1)}\frac{\Gamma(-2l)}{\Gamma(-l)} 
= \frac{2^{-2l-1}}{(-2l-1)\sqrt{\pi}}\Gamma\left(\frac{-2l+1}{2}\right)\\
&= \frac{2^{-2l-1}}{(-2l-1)\sqrt{\pi}} \frac{(-1)^l 2^{2l-1}\sqrt{\pi}(l-1)!}{(2l-1)!}\\
&= \frac{(-1)^{l+1}l!}{2(2l+1)!}.
\end{split}
\end{align}

\begin{acknowledgements}
We thank V. Cardoso for the beneficial comments. This work is supported in part by the National Nature Science Foundation of China (NSFC) under grants No.~12075136 and the Natural Science Foundation of Shandong Province under grant No.~ZR2020MA094. HZ was also supported by the Alexander von Humboldt Foundation in the early stage of this work.
\end{acknowledgements}



\begin{thebibliography}{99}

\bibitem{Detweiler:1980uk}
S.~L.~Detweiler,
Phys. Rev. D \textbf{22}, 2323-2326 (1980)

\bibitem{Pani:2012bp}
P.~Pani, V.~Cardoso, L.~Gualtieri, E.~Berti and A.~Ishibashi,
Phys. Rev. D \textbf{86}, 104017 (2012)
[arXiv:1209.0773 [gr-qc]].

\bibitem{Huang:2018qdl}
Y.~Huang, D.~J.~Liu, X.~h.~Zhai and X.~z.~Li,
Phys. Rev. D \textbf{98}, no.2, 025021 (2018)
[arXiv:1807.06263 [gr-qc]].

\bibitem{Boskovic:2018lkj}
M.~Boskovic, R.~Brito, V.~Cardoso, T.~Ikeda and H.~Witek,
Phys. Rev. D \textbf{99}, no.3, 035006 (2019)
[arXiv:1811.04945 [gr-qc]].

\bibitem{Franzin:2021kvj}
E.~Franzin, S.~Liberati and M.~Oi,
Phys. Rev. D \textbf{103}, no.10, 104034 (2021)
[arXiv:2102.03152 [gr-qc]].

\bibitem{Garcia-Saenz:2021uyv}
S.~Garcia-Saenz, A.~Held and J.~Zhang,
Phys. Rev. Lett. \textbf{127}, no.13, 131104 (2021)
[arXiv:2104.08049 [gr-qc]].

\bibitem{Lin:2021ssw}
J.~M.~Lin, M.~J.~Luo, Z.~H.~Zheng, L.~Yin and J.~H.~Huang,
Phys. Lett. B \textbf{819}, 136392 (2021)
[arXiv:2105.02161 [gr-qc]].

\bibitem{Biswas:2021gvq}
S.~Biswas,
Phys. Lett. B \textbf{820}, 136597 (2021)
[arXiv:2106.13837 [gr-qc]].

\bibitem{Guo:2021xao}
R.~Z.~Guo, C.~Yuan and Q.~G.~Huang,
Phys. Rev. D \textbf{105}, no.6, 064029 (2022)
[arXiv:2109.03376 [gr-qc]].

\bibitem{Herdeiro:2021znw}
C.~A.~R.~Herdeiro, E.~Radu and N.~M.~Santos,
Phys. Lett. B \textbf{824}, 136835 (2022)
[arXiv:2111.03667 [gr-qc]].

\bibitem{Arvanitaki:2010sy}
A.~Arvanitaki and S.~Dubovsky,
Phys. Rev. D \textbf{83}, 044026 (2011)
[arXiv:1004.3558 [hep-th]].

\bibitem{Hui:2021tkt}
L.~Hui,
Ann. Rev. Astron. Astrophys. \textbf{59}, 247-289 (2021)
[arXiv:2101.11735 [astro-ph.CO]].

\bibitem{Roy:2019esk}
R.~Roy and U.~A.~Yajnik,
Phys. Lett. B \textbf{803}, 135284 (2020)
[arXiv:1906.03190 [gr-qc]].

\bibitem{Creci:2020mfg}
G.~Creci, S.~Vandoren and H.~Witek,
Phys. Rev. D \textbf{101}, no.12, 124051 (2020)
[arXiv:2004.05178 [gr-qc]].

\bibitem{Davoudiasl:2019nlo}
H.~Davoudiasl and P.~B.~Denton,
Phys. Rev. Lett. \textbf{123}, no.2, 021102 (2019)
[arXiv:1904.09242 [astro-ph.CO]].

\bibitem{Cunha:2019ikd}
P.~V.~P.~Cunha, C.~A.~R.~Herdeiro and E.~Radu,
Universe \textbf{5}, no.12, 220 (2019)
[arXiv:1909.08039 [gr-qc]].

\bibitem{Roy:2021uye}
R.~Roy, S.~Vagnozzi and L.~Visinelli,
Phys. Rev. D \textbf{105}, no.8, 083002 (2022)
[arXiv:2112.06932 [astro-ph.HE]].

\bibitem{Yuan:2020xui}
G.~W.~Yuan, Z.~Q.~Xia, C.~Tang, Y.~Zhao, Y.~F.~Cai, Y.~Chen, J.~Shu and Q.~Yuan,
JCAP \textbf{03}, 018 (2021)
[arXiv:2008.13662 [astro-ph.HE]].

\bibitem{Chen:2019fsq}
Y.~Chen, J.~Shu, X.~Xue, Q.~Yuan and Y.~Zhao,
Phys. Rev. Lett. \textbf{124}, no.6, 061102 (2020)
[arXiv:1905.02213 [hep-ph]].

\bibitem{Cannizzaro:2021zbp}
E.~Cannizzaro, A.~Caputo, L.~Sberna and P.~Pani,
Phys. Rev. D \textbf{104}, no.10, 104048 (2021)
[arXiv:2107.01174 [gr-qc]].

\bibitem{Liu:2021llm}
T.~Liu and K.~F.~Lyu,
[arXiv:2107.09971 [astro-ph.HE]].

\bibitem{Arvanitaki:2014wva}
A.~Arvanitaki, M.~Baryakhtar and X.~Huang,
Phys. Rev. D \textbf{91}, no.8, 084011 (2015)
[arXiv:1411.2263 [hep-ph]].

\bibitem{Arvanitaki:2016qwi}
A.~Arvanitaki, M.~Baryakhtar, S.~Dimopoulos, S.~Dubovsky and R.~Lasenby,
Phys. Rev. D \textbf{95}, no.4, 043001 (2017)
[arXiv:1604.03958 [hep-ph]].

\bibitem{Baryakhtar:2017ngi}
M.~Baryakhtar, R.~Lasenby and M.~Teo,
Phys. Rev. D \textbf{96}, no.3, 035019 (2017)
[arXiv:1704.05081 [hep-ph]].

\bibitem{Brito:2017wnc}
R.~Brito, S.~Ghosh, E.~Barausse, E.~Berti, V.~Cardoso, I.~Dvorkin, A.~Klein and P.~Pani,
Phys. Rev. Lett. \textbf{119}, no.13, 131101 (2017)
[arXiv:1706.05097 [gr-qc]].

\bibitem{Brito:2017zvb}
R.~Brito, S.~Ghosh, E.~Barausse, E.~Berti, V.~Cardoso, I.~Dvorkin, A.~Klein and P.~Pani,
Phys. Rev. D \textbf{96}, no.6, 064050 (2017)
[arXiv:1706.06311 [gr-qc]].

\bibitem{Hannuksela:2018izj}
O.~A.~Hannuksela, K.~W.~K.~Wong, R.~Brito, E.~Berti and T.~G.~F.~Li,
Nature Astron. \textbf{3}, no.5, 447-451 (2019)
[arXiv:1804.09659 [astro-ph.HE]].

\bibitem{Isi:2018pzk}
M.~Isi, L.~Sun, R.~Brito and A.~Melatos,
Phys. Rev. D \textbf{99}, no.8, 084042 (2019)
[erratum: Phys. Rev. D \textbf{102}, no.4, 049901 (2020)]
[arXiv:1810.03812 [gr-qc]].

\bibitem{Palomba:2019vxe}
C.~Palomba, S.~D'Antonio, P.~Astone, S.~Frasca, G.~Intini, I.~La Rosa, P.~Leaci, S.~Mastrogiovanni, A.~L.~Miller and F.~Muciaccia, \textit{et al.}
Phys. Rev. Lett. \textbf{123}, 171101 (2019)
[arXiv:1909.08854 [astro-ph.HE]].

\bibitem{Sun:2019mqb}
L.~Sun, R.~Brito and M.~Isi,
Phys. Rev. D \textbf{101}, no.6, 063020 (2020)
[erratum: Phys. Rev. D \textbf{102}, no.8, 089902 (2020)]
[arXiv:1909.11267 [gr-qc]].

\bibitem{Brito:2020lup}
R.~Brito, S.~Grillo and P.~Pani,
Phys. Rev. Lett. \textbf{124}, no.21, 211101 (2020)
[arXiv:2002.04055 [gr-qc]].

\bibitem{Barausse:2020rsu}
E.~Barausse, E.~Berti, T.~Hertog, S.~A.~Hughes, P.~Jetzer, P.~Pani, T.~P.~Sotiriou, N.~Tamanini, H.~Witek and K.~Yagi, \textit{et al.}
Gen. Rel. Grav. \textbf{52}, no.8, 81 (2020)
[arXiv:2001.09793 [gr-qc]].

\bibitem{Zhu:2020tht}
S.~J.~Zhu, M.~Baryakhtar, M.~A.~Papa, D.~Tsuna, N.~Kawanaka and H.~B.~Eggenstein,
Phys. Rev. D \textbf{102}, no.6, 063020 (2020)
[arXiv:2003.03359 [gr-qc]].

\bibitem{Ng:2020jqd}
K.~K.~Y.~Ng, M.~Isi, C.~J.~Haster and S.~Vitale,
Phys. Rev. D \textbf{102}, no.8, 083020 (2020)
[arXiv:2007.12793 [gr-qc]].

\bibitem{Baryakhtar:2020gao}
M.~Baryakhtar, M.~Galanis, R.~Lasenby and O.~Simon,
Phys. Rev. D \textbf{103}, no.9, 095019 (2021)
[arXiv:2011.11646 [hep-ph]].

\bibitem{Aggarwal:2020olq}
N.~Aggarwal, O.~D.~Aguiar, A.~Bauswein, G.~Cella, S.~Clesse, A.~M.~Cruise, V.~Domcke, D.~G.~Figueroa, A.~Geraci and M.~Goryachev, \textit{et al.}
Living Rev. Rel. \textbf{24}, no.1, 4 (2021)
[arXiv:2011.12414 [gr-qc]].

\bibitem{Chen:2021lvo}
Y.~Chen, Y.~Liu, R.~S.~Lu, Y.~Mizuno, J.~Shu, X.~Xue, Q.~Yuan and Y.~Zhao,
Nature Astron. \textbf{6}, no.5, 592-598 (2022)
[arXiv:2105.04572 [hep-ph]].

\bibitem{Yuan:2021ebu}
C.~Yuan, R.~Brito and V.~Cardoso,
Phys. Rev. D \textbf{104}, no.4, 044011 (2021)
[arXiv:2106.00021 [gr-qc]].

\bibitem{Ghosh:2021uqw}
S.~Ghosh,
Mod. Phys. Lett. A \textbf{36}, no.33, 2130024 (2021)
[arXiv:2111.09394 [gr-qc]].

\bibitem{Berlin:2021txa}
A.~Berlin, D.~Blas, R.~Tito D'Agnolo, S.~A.~R.~Ellis, R.~Harnik, Y.~Kahn and J.~Sch\"utte-Engel,
Phys. Rev. D \textbf{105}, no.11, 116011 (2022)
[arXiv:2112.11465 [hep-ph]].

\bibitem{Ghosh:2018gaw}
S.~Ghosh, E.~Berti, R.~Brito and M.~Richartz,
Phys. Rev. D \textbf{99}, no.10, 104030 (2019)
[arXiv:1812.01620 [gr-qc]].

\bibitem{Baumann:2018vus}
D.~Baumann, H.~S.~Chia and R.~A.~Porto,
Phys. Rev. D \textbf{99}, no.4, 044001 (2019)
[arXiv:1804.03208 [gr-qc]].

\bibitem{Zhang:2019eid}
J.~Zhang and H.~Yang,
Phys. Rev. D \textbf{101}, no.4, 043020 (2020)
[arXiv:1907.13582 [gr-qc]].

\bibitem{Hannuksela:2019vip}
O.~A.~Hannuksela, K.~C.~Y.~Ng and T.~G.~F.~Li,
Phys. Rev. D \textbf{102}, no.10, 103022 (2020)
[arXiv:1906.11845 [astro-ph.CO]].

\bibitem{Annulli:2020lyc}
L.~Annulli, V.~Cardoso and R.~Vicente,
Phys. Rev. D \textbf{102}, no.6, 063022 (2020)
[arXiv:2009.00012 [gr-qc]].

\bibitem{Ikeda:2020xvt}
T.~Ikeda, L.~Bernard, V.~Cardoso and M.~Zilh\~ao,
Phys. Rev. D \textbf{103}, no.2, 024020 (2021)
[arXiv:2010.00008 [gr-qc]].

\bibitem{Choudhary:2020pxy}
S.~Choudhary, N.~Sanchis-Gual, A.~Gupta, J.~C.~Degollado, S.~Bose and J.~A.~Font,
Phys. Rev. D \textbf{103}, no.4, 044032 (2021)
[arXiv:2010.00935 [gr-qc]].

\bibitem{Tong:2021whq}
X.~Tong, Y.~Wang and H.~Y.~Zhu,
Astrophys. J. \textbf{924}, no.2, 99 (2022)
[arXiv:2106.13484 [astro-ph.HE]].

\bibitem{DeLuca:2021ite}
V.~De Luca and P.~Pani,
JCAP \textbf{08}, 032 (2021)
[arXiv:2106.14428 [gr-qc]].

\bibitem{Chung:2021roh}
A.~K.~W.~Chung, J.~Gais, M.~H.~Y.~Cheung and T.~G.~F.~Li,
Phys. Rev. D \textbf{104}, no.8, 084028 (2021)
[arXiv:2107.05492 [gr-qc]].

\bibitem{Payne:2021ahy}
E.~Payne, L.~Sun, K.~Kremer, P.~D.~Lasky and E.~Thrane,
Astrophys. J. \textbf{931}, no.2, 79 (2022)
[arXiv:2107.11730 [gr-qc]].

\bibitem{Su:2021dwz}
B.~Su, Z.~Z.~Xianyu and X.~Zhang,
Astrophys. J. \textbf{923}, no.1, 114 (2021)
[arXiv:2107.13527 [gr-qc]].

\bibitem{Takahashi:2021yhy}
T.~Takahashi, H.~Omiya and T.~Tanaka,
PTEP \textbf{2022}, no.4, 043E01 (2022)
[arXiv:2112.05774 [gr-qc]].

\bibitem{Brito:2015oca}
R.~Brito, V.~Cardoso and P.~Pani,
Physics,''
Lect. Notes Phys. \textbf{906}, pp.1-237 (2015)
[arXiv:1501.06570 [gr-qc]].

\bibitem{Vishveshwara:1970zz}
C.~V.~Vishveshwara,
Nature \textbf{227}, 936-938 (1970)

\bibitem{Leaver:1985ax}
E.~W.~Leaver,
Proc. Roy. Soc. Lond. A \textbf{402}, 285-298 (1985)

\bibitem{Cardoso:2005vk}
V.~Cardoso and S.~Yoshida,
JHEP \textbf{07}, 009 (2005)
[arXiv:hep-th/0502206 [hep-th]].

\bibitem{Dolan:2007mj}
S.~R.~Dolan,
Phys. Rev. D \textbf{76}, 084001 (2007)
[arXiv:0705.2880 [gr-qc]].

\bibitem{Zouros:1979iw}
T.~J.~M.~Zouros and D.~M.~Eardley,
Annals Phys. \textbf{118}, 139-155 (1979)

\bibitem{Boyer:1966qh}
R.~H.~Boyer and R.~W.~Lindquist,
J. Math. Phys. \textbf{8}, 265 (1967)

\bibitem{Press:1973zz}
W.~H.~Press and S.~A.~Teukolsky,
Astrophys. J. \textbf{185}, 649-674 (1973)

\bibitem{Berti:2005gp}
E.~Berti, V.~Cardoso and M.~Casals,
Phys. Rev. D \textbf{73}, 024013 (2006)
[erratum: Phys. Rev. D \textbf{73}, 109902 (2006)]
[arXiv:gr-qc/0511111 [gr-qc]].

\end{thebibliography}
\end{document}